\documentclass[conference]{IEEEtran}
\IEEEoverridecommandlockouts
\usepackage{cite}
\usepackage{amsmath,amssymb,amsfonts}
\usepackage{algorithmic}
\usepackage{graphicx}
\usepackage{textcomp}
\usepackage{xcolor}
\usepackage{url}
\usepackage{hyperref}

\usepackage[colorinlistoftodos]{todonotes}
\usepackage[normalem]{ulem}

\usepackage{amsmath}

\def\BibTeX{{\rm B\kern-.05em{\sc i\kern-.025em b}\kern-.08em
    T\kern-.1667em\lower.7ex\hbox{E}\kern-.125emX}}
\begin{document}

\title{A Catalog of Micro Frontends Anti-patterns}

\author{\IEEEauthorblockN{1\textsuperscript{st} Nabson Paiva Souza da Silva}
\IEEEauthorblockA{\textit{Federal University of Amazonas}\\
Manaus, AM, Brazil \\
nabson.paiva@icomp.ufam.edu.br}
\and
\IEEEauthorblockN{2\textsuperscript{nd} Eriky Rodrigues}
\IEEEauthorblockA{\textit{Federal University of Amazonas}\\
Manaus, AM, Brazil \\
eriky.rodrigues@icomp.ufam.edu.br}
\and
\IEEEauthorblockN{3\textsuperscript{th} Tayana Conte}
\IEEEauthorblockA{\textit{Federal University of Amazonas}\\
Manaus, AM, Brazil \\
tayana@icomp.ufam.edu.br}
}
\maketitle

\begin{abstract}
Micro frontend (MFE) architectures have gained significant popularity for promoting independence and modularity in development. Despite their widespread adoption, the field remains relatively unexplored, especially concerning identifying problems and documenting best practices. Drawing on both established microservice (MS) anti-patterns and the analysis of real problems faced by software development teams that adopt MFE, this paper presents a catalog of 12 MFE anti-patterns. We composed an initial version of the catalog by recognizing parallels between MS anti-patterns and recurring issues in MFE projects to map and adapt MS anti-patterns to the context of MFE. To validate the identified problems and proposed solutions, we conducted a survey with industry practitioners, collecting valuable feedback to refine the anti-patterns. Additionally, we asked participants if they had encountered these problems in practice and to rate their harmfulness on a 10-point Likert scale. The survey results revealed that participants had encountered all the proposed anti-patterns in real-world MFE architectures, with only one reported by less than 50\% of participants. They stated that the catalog can serve as a valuable guide for both new and experienced developers, with the potential to enhance MFE development quality. The collected feedback led to the development of an improved version of the anti-patterns catalog. Furthermore, we developed a web application designed to not only showcase the anti-patterns but also to actively foster collaboration and engagement within the MFE community. The proposed catalog is a valuable resource for identifying and mitigating potential pitfalls in MFE development. It empowers developers of all experience levels to create more robust, maintainable, and well-designed MFE applications.
\end{abstract}


\section{Introduction}

As a monolithic application grows, it becomes challenging to scale due to limitations like technology constraints, the necessity for vertical scaling only, and the need to reboot the entire application with each deployment~\cite{dragoni2017microservices}. To address these issues, developers are adopting the microservice architectural style to create autonomous, distributed, and loosely coupled services~\cite{dmitry2014micro,lewis2014microservices,erl2016service}. This architecture enables teams to work independently, reducing the development time for new features. However, different teams often still need to share the same codebase for the presentation layer.

Micro frontends offer a solution by extending the principles of microservices to the frontend, thereby enabling independent testing, development, and deployment of frontend components~\cite{geers2020micro,mezzalira2021building,peltonen2021motivations}. This architectural style breaks down a frontend application into smaller, manageable slices. Many companies, such as SAP, Springer, Zalando, NewRelic, Ikea, Starbucks, and DAZN, have successfully adopted the micro frontend architecture~\cite{taibi2022micro}.

Over time, software architecture can deteriorate due to developers' insufficient understanding of the specific architectural style~\cite{taibi2020microservices}. This issue is particularly critical in micro frontend architecture, as there is no well-defined method for evaluating this type of architecture. Therefore, we propose a catalog of 12 anti-patterns for micro frontends to preserve architectural integrity and assist developers in making well-informed decisions. Anti-patterns can address emerging issues, common mistakes, poorly implemented solutions, misapplied best practices, and deviations from established process models~\cite{brada2019software}.

The anti-patterns were defined based on microservices anti-patterns, given their similarity to micro frontends, and the analysis of real problems faced by software development teams that adopt micro frontends. Each anti-pattern includes a practical example to aid in understanding the associated problem. To evaluate the issues and solutions associated with each anti-pattern, we conducted a survey to gather feedback from 20 industry practitioners with experience in micro frontend development. We assessed the severity of each anti-pattern on a 10-point Likert Scale~\cite{likert1932technique}, ranging from ``Not harmful" to ``Extremely harmful". Following a Code Reliability Thematic Analysis~\cite{braun2021one} of the survey's qualitative responses, we refined the catalog to generate its improved version, incorporating the practitioners' feedback. We used Cohen's Kappa~\cite{cohen1960coefficient} as a measure of agreement between two coders to ensure coding reliability.

The survey results revealed that all anti-patterns have been observed in real-world architectures, with only one reported by less than 50\% of industry practitioners. Our findings indicate that 8 out of 12 anti-patterns have a median harmfulness score above 8, including 10 and 9 scores. Through qualitative feedback, we identified that participants emphasized that the catalog serves as a guide for both new and experienced developers. They highlighted its ability to address numerous real-world problems encountered in daily work with micro frontends, underscoring its potential to enhance development quality significantly.  Based on participant feedback to improve the problems and solutions associated with each anti-pattern, we generated an improved version of the catalog, which will be presented in this paper. Additionally, we developed a web application to showcase all anti-patterns, allowing practitioners to collaborate on the catalog and propose new anti-patterns.


The key contributions of this paper are (1) a catalog of micro frontend anti-patterns inspired by microservices anti-patterns; (2) an empirical validation of these anti-patterns with industry practitioners; and (3) a collaborative micro frontends anti-patterns repository to facilitate knowledge-sharing within the software development community. By bridging the gap between academic research and industry practice, this work aims to empower developers with actionable insights to avoid common pitfalls and enhance the overall quality of micro frontend applications.

\section{Background}
This section delves into the fundamental principles of microservices, micro frontends, and anti-patterns. It provides readers with the necessary knowledge to appreciate the context and significance of our work.

\subsection{Anti-patterns}
Anti-patterns are recurring design practices, choices, or solutions to common problems. Despite appearing reasonable and effective, they lead to negative consequences and undermine the system’s overall quality~\cite{cerny2023catalog}. An anti-pattern is similar to a pattern, but instead of providing a practical solution, it offers an approach that appears to be a solution on the surface. However, it ultimately leads to more problems~\cite{koenig1998patterns}. While a problem and its optimal solution characterize patterns, anti-patterns involve two solutions~\cite{brown1998antipatterns}. The first solution is a commonly occurring solution that generates overwhelmingly negative consequences. The second solution is a commonly occurring method in which the anti-pattern can be resolved and reengineered into a more beneficial form.

There are three forms to define anti-patterns~\cite{koenig1998patterns,brada2019software}: (1) Degenerative Form: the most basic one is a textual description without any structure, template, or separate content sections for various aspects of the pattern; (2) Mini-AntiPattern: a more structured approach, consists of name, problem, and solution; and (3) Formal Definitions: including the Full AntiPattern Template and the Laplante-Neil Structure, consists of multiple fields detailing various dimensions of the anti-pattern. In this paper, we will use the Mini-AntiPattern format due to its simplicity in proposing new anti-patterns while maintaining a structured approach.

\subsection{Microservices}
Microservice is an architectural style presented as an alternative to monolith architectures~\cite{abgaz2023decomposition}. Lewis and Fowler~\cite{lewis2014microservices} first defined Microservice as an approach to developing a single application as a suite of small services, each running in its process and communicating with lightweight mechanisms. A service is a self-contained, loosely coupled software unit designed to perform a specific business function. This modular design ensures each service has a well-defined and focused set of responsibilities, promoting maintainability and scalability~\cite{erl2016service}. Microservices enable technology heterogeneity, autonomous teams, independent deployment, independent scale, independent testing, and easy experimenting and adoption of new technologies~\cite{richardson2018microservices,newman2021building}.

Microservices anti-patterns are well-defined in scientific literature. Some of them can be correlated with corresponding micro frontends anti-patterns, owing the shared characteristics between both architectural styles: Cyclic Dependency~\cite{taibi2020microservices,tighilt2020study,parker2023visualizing,cerny2023catalog,bogner2019towards}, Hub-like Dependency~\cite{cerny2023catalog}, The Knot or Knot Service~\cite{parker2023visualizing,cerny2023catalog}, Microservice Greedy~\cite{taibi2020microservices,parker2023visualizing,cerny2023catalog}, Nano Service~\cite{tighilt2020study,parker2023visualizing,cerny2023catalog,bogner2019towards}, Mega service~\cite{taibi2020microservices,tighilt2020study,cerny2023catalog,bogner2019towards}, Wrong Cuts~\cite{taibi2020microservices,parker2023visualizing,cerny2023catalog,bogner2019towards}, No Api Versioning~\cite{taibi2020microservices,tighilt2020study,cerny2023catalog,bogner2019towards}, No DevOps Tools or Continuous Integration (CI) / Continuous Delivery (CD)~\cite{taibi2020microservices,tighilt2020study,cerny2023catalog}, Microservices as the goal~\cite{taibi2020microservices}, Lack of Microservice Skeleton~\cite{taibi2020microservices} and Golden hammer~\cite{bogner2019towards}. These anti-patterns formed the basis for proposing micro frontend anti-patterns.

\subsection{Micro frontends}
The term ``Micro Frontend'' was first coined by ThoughtWorks in 2016~\cite{thoughtworks2023microfrontends} as an architectural style inspired by microservices architecture. The main idea is to decompose a monolithic frontend application into smaller parts that can be developed, deployed, and updated independently, promoting greater flexibility and maintainability~\cite{mezzalira2021building,peltonen2021motivations}. Micro frontend (MFE) can also be considered an organizational approach. The application is divided into vertical slices built from the database to the user interface and run by a dedicated team~\cite{geers2020micro}. In an MFE architecture, the web application integrates different features or business sub-domains, and each software team should have only one domain to handle~\cite{peltonen2021motivations}. Many companies adopted the MFE architecture, such as SAP, Springer, Zalando, NewRelic, Ikea, Starbucks, and DAZN~\cite{taibi2022micro}.

MFEs share the main principles, benefits, and issues of microservices~\cite{thoughtworks2023microfrontends,taibi2022micro,mezzalira2021building}. The motivations are development scalability and codebase growth, and the benefits include support for different technologies, autonomous cross-functional teams, development, deployment, management independence, and better testability. However, the primary drawbacks are increased payload size, complex monitoring and debugging, state management, and duplicated code~\cite{peltonen2021motivations}.

Each MFE implements a set of screens and fragments. Fragments are reusable User Interface (UI) components that can be combined to form screens across different MFEs~\cite{geers2020micro}. Some fragments might need context information, but the team, including the fragment in their MFE, does not have to know the fragments' state or implementation details. The MFEs must integrate a coherent application to deliver a unified User Experience (UX)~\cite{geers2020micro}. Achieving this requires developers to make informed decisions regarding the composition, communication, and routing between the MFEs.

The first decision is about composition, the process of requesting the fragments and putting them in the correct slots in a screen~\cite{geers2020micro}. There are three approaches to composing MFEs: Server-side, Edge-side, and Client-side.

\begin{enumerate}
    \item \textbf{Client-side Composition (CSC)}: An application shell loads MicroFrontends inside itself. An application shell is technically represented by an HTML file always present during the user session containing a small JavaScript code for loading and orchestrating different MFEs~\cite{taibi2022micro,peltonen2021motivations,geers2020micro}. CSC needs MFEs-specific frameworks~\cite{moraes2024micro}, such as single-spa~\cite{single2016single}, qiankun~\cite{qiankun2019qiankun} and Garfish~\cite{garfish2021garfish}, or with frameworkless technologies like Webpack, Module Federation, iFrames, and web components.
    \item \textbf{Edge-side Composition (ESC)}: The web page is assembled at the Content Delivery Network (CDN) level, using an XML-based markup language called Edge Side Include (ESI)~\cite{peltonen2021motivations}. One of the drawbacks of this implementation is that ESI is not implemented in the same way by each CDN provider, which can lead to many refactors and new logic implementation.
    \item \textbf{Server-side Composition (SSC)}: The origin server is composing the view by retrieving all the different MFEs and assembling the final page. It can happen at runtime or compile time~\cite{peltonen2021motivations,geers2020micro}. It is the simplest approach because it enables the development of MFEs as packages~\cite{moraes2024micro}. However, this approach does not meet some of the main principles of MFE, such as technology agnosticism and independent delivery.
\end{enumerate}

Once the fragments are composed into a screen, the team needs to decide how the MFEs will communicate with each other. Effective communication outlines how the screen and fragments interact to deliver a seamless user experience (UX)~\cite{taibi2022micro}. Geers~\cite{geers2020micro} defines three primary forms of communication: Parent to Fragment, Fragment to Parent, and Fragment to Fragment.

\begin{enumerate}
    \item \textbf{Parent to Fragment Communication}: A change in the screen is propagated down to one or more fragments so they can update themselves. This form is also called Parent-child Communication.
    \item \textbf{Fragment to Parent Communication}: A change on a fragment sends a message to the screen so it can update itself.  This form is also called Child-parent Communication.
    \item \textbf{Fragment to Fragment Communication}: A change on a fragment sends a message to one or more fragments composed on the same screen. This form is also called Child-child Communication.
\end{enumerate}

The final decision involves routing, which delineates the navigation from one view to another~\cite{taibi2022micro}. Usually, hyperlinks are necessary to navigate between screens. When adopting CSC, the application shell manages routing, which knows all routes and MFEs, mapping URLs to the correct MFE.

\section{Methodology}

\begin{figure}[htbp]
\centerline{\includegraphics[width=1\linewidth]{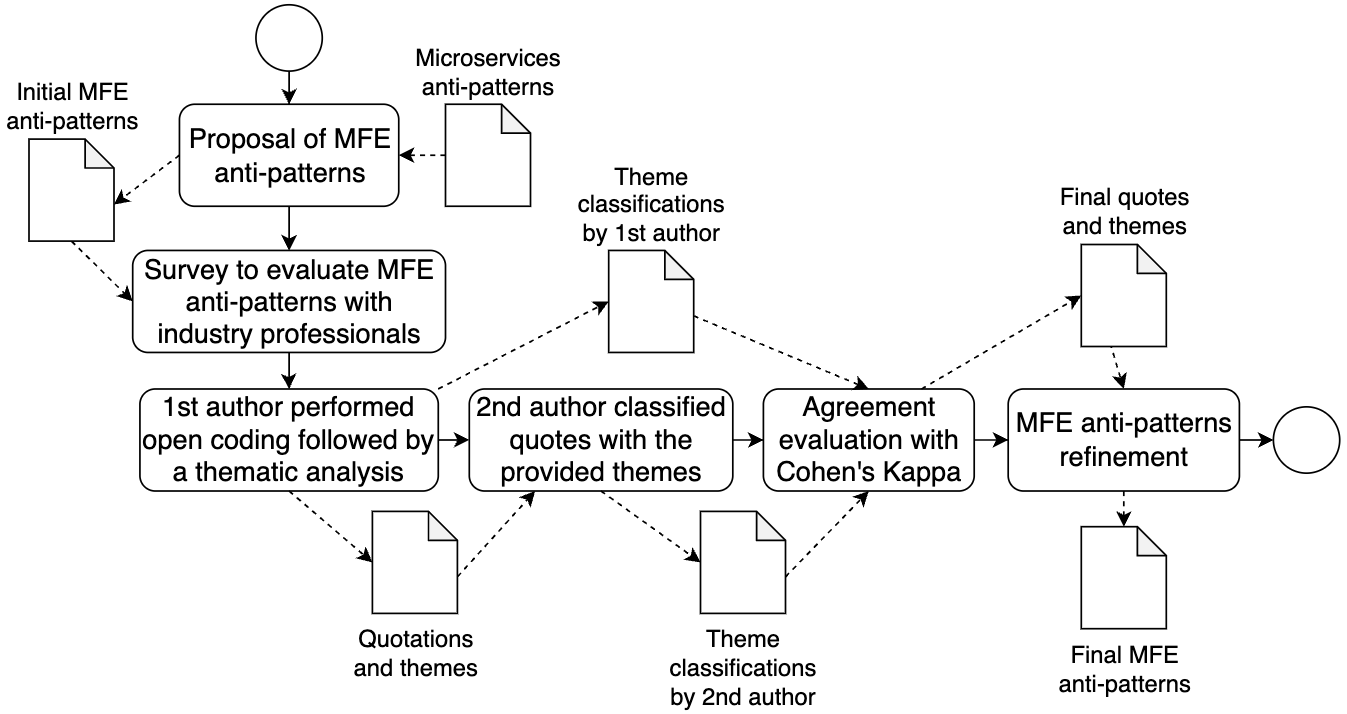}}
\caption{BPMN diagram illustrating the process we followed to propose and refine the MFE anti-patterns.}
\label{fig:method}
\end{figure}

Figure~\ref{fig:method} illustrates the process of proposing and refining the MFE anti-patterns, which we detail in this Section. Initially, we used the Systematic Literature Review (SLR) of Cerny et al.~\cite{cerny2023catalog} as a basis for identifying microservices anti-patterns. This SLR provided a comprehensive catalog of 58 disjoint anti-patterns derived from 203 originally identified anti-patterns for microservices. Drawing upon the microservice anti-patterns identified by Cerny et al. and further explored in Taibi et al.~\cite{taibi2020microservices}, Tighilt et al.~\cite{tighilt2020study}, Parker et al.~\cite{parker2023visualizing} and Bogner et al.~\cite{bogner2019towards}, coupled with an analysis of real-world problems faced by MFE development teams, we proposed a catalog of 12 MFE anti-patterns. Each anti-pattern follows the Mini-AntiPattern format (name, problem, and solution) and includes an example to illustrate the problem within an MFE context. The initial proposal of the anti-patterns can be found in our supplementary material (Section~\ref{sec:data}). We classified the anti-patterns into four categories:

\begin{enumerate}
    \item \textbf{Intra-fronted category}: considers a single MFE component and its design.
    \item \textbf{Inter-frontend category}: considers the structural division and communication involving two or more MFEs.
    \item \textbf{Operations category}: related to the operational practices and continuous maintenance of the application.
    \item \textbf{Development category}: related to the development team and their decisions around the architecture.
\end{enumerate}

To evaluate the proposed anti-patterns, we conducted a survey that included both open and closed questions. The intended audience for the survey included software industry practitioners with experience in MFE development. This approach allowed the validation of different aspects of the anti-patterns by incorporating the insights of practitioners specialized in the MFE field. We also included a consent form as part of our survey instruments, allowing participants to provide their consent to participate. We assured participants that their participation was voluntary and that they could withdraw at any time. All collected data was anonymized, ensuring privacy and enabling participants to share valuable insights without concern. Our survey comprised three sections and can be found in our supplementary material (Section~\ref{sec:data}).

The survey's first section aims to understand the level of practitioner experience and the role of participants in the context of MFE development, ensuring that the feedback provided comes from practitioners with practical knowledge in this area. Additionally, the questions allowed the characterization of each participant. The questions focused on determining whether the practitioners currently work with MFE, their experience in this area quantified in years, and their role in MFE-related projects.



The survey's second section aimed to validate each of the 12 anti-patterns proposed in the catalog. This involved analyzing each anti-pattern through a series of questions designed to: (1) verify whether the descriptions of the problems and proposed solutions were clear and understandable to participants; (2) determine if the proposed solution effectively addresses the problem described in the anti-pattern, and if not, gather alternative solutions; (3) identify the practical occurrence of the anti-pattern in participants' projects; and (4) assess participants' perception of the anti-pattern's impact on MFE projects by assigning a harmfulness value. To define the harmfulness value of each anti-pattern, we used a 10-point Likert scale based on Taibi et al.~\cite{taibi2020microservices}, where 1 means ``Not harmful" and 10 indicates ``Extremely harmful". Each page of this section included the following questions to evaluate each anti-pattern individually:

\begin{itemize}
    \item Is the anti-pattern problem clearly stated?  
    \item If you disagree, please provide a description of what is not clear
    \item Is the anti-pattern solution clearly stated?  
    \item If you disagree, please provide a description of what is not clear
    \item Does the proposed solution address the problem presented in the anti-pattern?  
    \item In case you have a different suggestion on the problem solution, please provide a description
    \item Have you ever encountered this problem in any project you have previously worked on or are currently working on?
    \item How harmful do you think this anti-pattern problem is?
\end{itemize}

The third section of the survey enabled participants to offer insights based on their practical experience, providing additional feedback and suggestions for improving the catalog. The questions were designed to identify any MFE-related issues not covered in the catalog, assess the potential impact of the catalog on the quality of MFE solutions, and gather recommendations for enhancing the catalog.

After collecting participants' responses to the survey, we conducted a quantitative analysis to identify the most common anti-patterns and calculated the median harmfulness score for each. To further investigate the harmfulness scores, we employed several statistical tests. First, we used the Shapiro-Wilk test~\cite{shapiro1965analysis} to determine whether the score samples for each anti-pattern were normally distributed. Since only half of the samples followed a normal distribution, we opted for non-parametric tests. As the same participants rated different anti-patterns, we treated the samples as dependent and selected the Friedman test~\cite{friedman1937use}, which is suitable for dependent samples. We applied the av post-hoc Dunn test~\cite{dunn1964multiple} to compare pairs of medians, as it is commonly used following Friedman test when significant differences are found.

For the qualitative feedback, we performed a thematic analysis grounded in coding reliability \cite{braun2021one}. Thematic analysis allows for organizing and categorizing feedback in a structured manner, which helps identify recurring patterns (themes). This approach aids in understanding participants' perspectives better and directing improvements based on the specific objectives of each piece of feedback. We assessed intercoder reliability (ICR) between two authors to ensure the robustness of the thematic analysis. The second author was included to evaluate the consistency and objectivity of the coding process. The first author conducted an initial open coding of all responses, resulting in the definition of eight themes. Subsequently, the second author independently applied these themes to the same quotations. To measure agreement between the authors, we calculated Cohen's Kappa \cite{cohen1960coefficient} and resolved any discrepancies through discussion to reach a consensus.

We used practitioners' feedback to refine the catalog and finalize the anti-patterns. Drawing on the collaborative repository model proposed by Bogner et al. \cite{bogner2019towards}, we developed a web application\footnote{\url{https://mfe-anti-patterns.online/micro-frontends-anti-patterns/\#/catalog}} to present all the anti-patterns. The application was built with ReactJS and is hosted on GitHub Pages, making it publicly accessible. Community members can contribute to the catalog by submitting Pull Requests (PR) to the Github repository\footnote{\url{https://github.com/nabsonp/micro-frontends-anti-patterns/pulls}} folowing the contribution guidelines outlined in the repository\footnote{\url{https://github.com/nabsonp/micro-frontends-anti-patterns/blob/main/CONTRIBUTING.md}}. All anti-patterns are stored in the \texttt{src/anti-patterns} folder\footnote{\url{https://github.com/nabsonp/micro-frontends-anti-patterns/blob/main/src/anti-patterns/index.ts}}, with each JSON file corresponding to a specific anti-pattern. Thanks to GitHub Actions, changes to the codebase are automatically reflected in the application when a PR is merged. 

\section{Results}

In this section, we present the analysis of the results obtained from the survey responses. We directly contacted a total of 37 practitioners to participate in the survey. However, due to the open-access nature of the survey link, precise control over participant recruitment was not feasible. As a consequence, we could not determine an exact response rate~\cite{ralph2020empirical}. A total of 20 practitioners volunteered to participate in the survey. Before responding, each participant signed a consent form, ensuring the confidentiality of the information.

\begin{table*}[]
\centering 
\label{tab:quantitative}
\caption{Overall results from quantitative analysis ranked by harmfulness score}
\begin{tabular}{|l|p{2cm}|p{2cm}|p{3cm}|p{1.7cm}|c|}
\hline
\textbf{Anti-pattern} & \textbf{Problem clearly stated rate} & \textbf{Solution clearly stated rate} & \textbf{Solution addresses the problem rate} & \textbf{AP seen in practice rate} & \textbf{Harmfulness} \\ \hline

No CI/CD & 100.00\% & 100.00\% & 100.00\% & 90.00\% & 10 \\ \hline

No Versioning & 100.00\% & 100.00\% & 95.00\%  & 70.00\%  & 9 \\ \hline

Common Ownership & 95.00\% & 95.00\% & 100.00\% & 55.00\% & 8 \\ \hline
Cyclic Dependency & 95.00\% & 95.00\%  & 100.00\% & 85.00\% & 8 \\ \hline
Hub-like Dependency & 95.00\% & 90.00\% & 95.00\% & 95.00\% & 8 \\ \hline
Knot Micro Frontend & 95.00\% & 95.00\% & 100.00\% & 80.00\% & 8 \\ \hline
Lack of Skeleton & 100.00\% & 100.00\% & 100.00\% & 65.00\% & 8 \\ \hline
Micro Frontend as the Goal & 100.00\% & 100.00\% & 100.00\% & 60.00\% & 8 \\ \hline

Mega Frontend & 100.00\% & 100.00\% & 100.00\% & 90.00\% & 7 \\ \hline
Micro Frontends Greedy & 95.00\% & 100.00\% & 100.00\% & 55.00\% & 7 \\ \hline
Nano Frontend & 100.00\% & 100.00\% & 100.00\% & 35.00\% & 7 \\ \hline

Golden Hammer & 95.00\% & 100.00\% & 100.00\% & 70.00\% & 6  \\ \hline

\end{tabular}
\end{table*}

On a brief overview of the participants' characterization, 75\% of the participants are currently involved in MFE software projects, and 75\% have more than two years of experience working with MFE as well. A total of 45\% of the subjects identified themselves as Fullstack Developers, 35\% as Frontend Developers, 10\% as Software Architects, and 5\% as Mobile Developers and SE Team Leaders each. We assigned each participant an identifier ranging from P1 to P20. More than half of the participants work at large companies with over 90 software engineers and projects involving 40+ MFEs and 70+ MS, with engineering teams structured as independent units. The remaining participants work on other software companies, but in smaller projects. This diversity ensures the identified anti-patterns are applicable across different scales and organizational structures. The participants' full characterization is available at our supplementary material (Section~\ref{sec:data}).

Subsection \ref{sec:quantitative} summarizes the quantitative results, whereas Subsection \ref{sec:qualitative} presents the thematic analysis findings identified from the collected qualitative data.

\subsection{Quantitative Analysis} \label{sec:quantitative}

The Table \ref{tab:quantitative} summarizes our quantitative results, which we ranked according to the harmful score values. Column 1 presents the titles of the anti-patterns evaluated in the survey. Columns 2 and 3 report the participants' agreement rates regarding the clarity of the problem presentation and the proposed solutions of the anti-patterns. Column 4 presents the participants' agreement rates regarding the effectiveness of the proposed solutions for the problems identified in the anti-patterns. Column 5 shows the percentage of participants who have encountered the problems described in the anti-patterns in their practitioner experience within the software industry. Column 6 presents the median values of harmfulness attributed to the anti-patterns by the participants.

Regarding the clarity of the problems and solutions presented in the anti-patterns, the results show that the participants had a thorough understanding, as demonstrated by the high values in Columns 2 and 3, which range from 95\% up to 100\%. Moreover, the high values presented in Column 3 indicate a positive efficacy of the solutions, on which a large majority of participants recognized the proposed solutions as effective means of addressing the issues posed by the anti-patterns.

The values in column 5 indicate that practitioners frequently encounter the anti-patterns Cyclic Dependency, Knot Frontend, Hub-like Dependency, Mega Frontend, and No CI/CD in software projects. Notably, seven of the remaining eight anti-patterns were reported by more than 50\% of participants, highlighting their common occurrence as well. The only anti-pattern observed by less than 50\% of participants is the Nano Frontend, in contrast to the Mega Frontend. This suggests that practitioners are more likely to create larger MFEs than smaller ones.

Given the limited sample size, we assessed the reliability of the harmfulness scores using several statistical tests. First, we applied the Shapiro-Wilk Test with a 95\% confidence level to determine if the samples were normally distributed. The test indicated that only half of the anti-patterns (namely Cyclic Dependency, Knot Micro Frontend, Nano Frontend, Mega Frontend and Golden Hammer) follow a normal distribution.

\begin{figure}[htbp]
\centerline{\includegraphics[width=\linewidth]{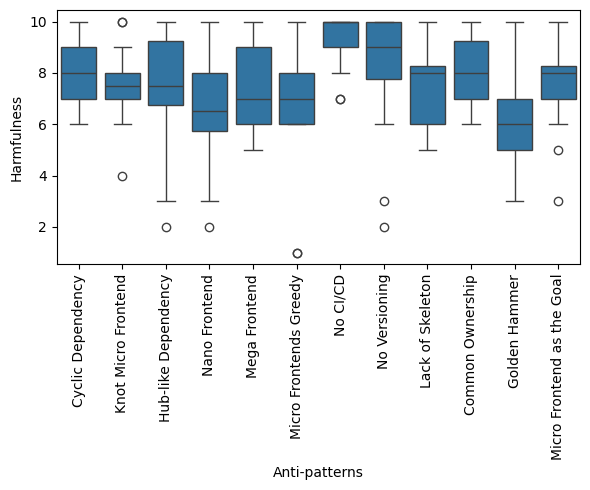}}
\caption{Boxplot illustrating the harmfulness ratings for each anti-pattern.}
\label{fig:boxplots}
\end{figure}

\begin{figure*}[ht]
    \centering
    \includegraphics[width=\textwidth]{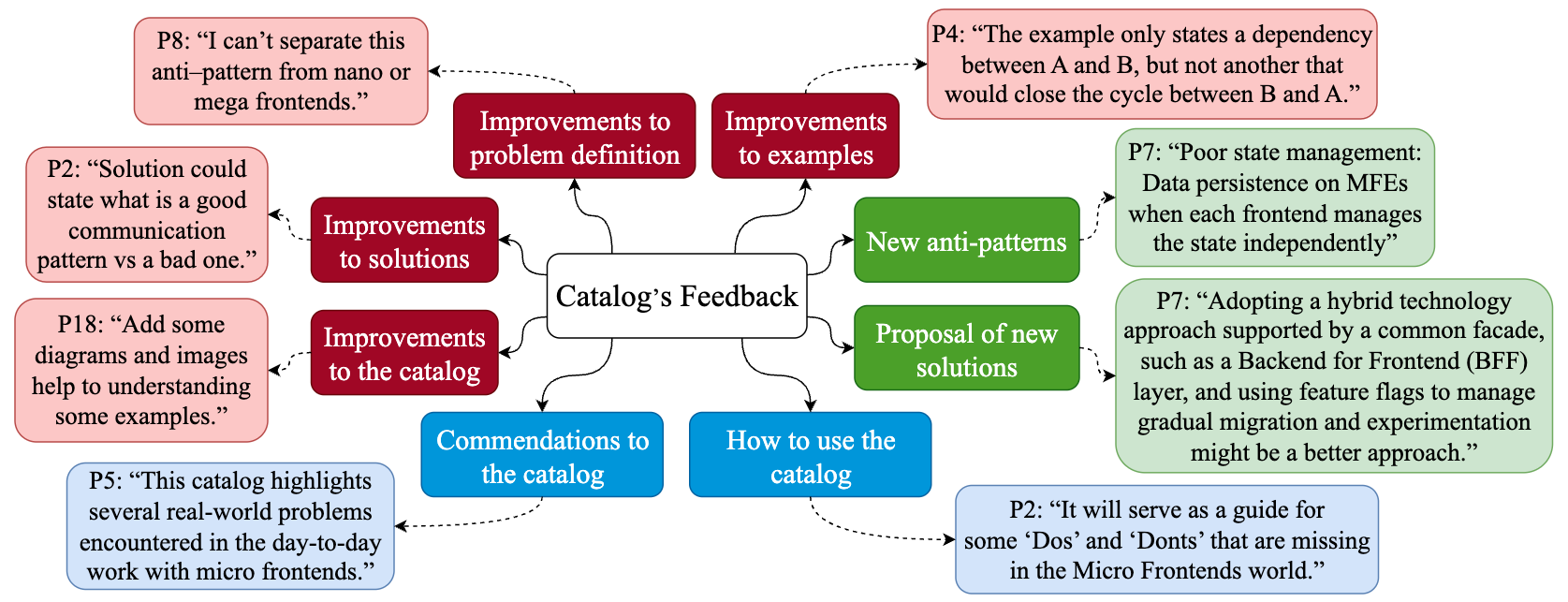}
    \caption{Identified themes during the Thematic Analysis, with accompanying example quotations. The red themes relate to suggested improvements, the green themes represent new proposals, and the blue themes highlight insights and recommendations for using the catalog.}
    \label{fig:themes}
\end{figure*}

We then used the Friedman test to evaluate whether there were statistically significant differences in the harmfulness scores. The Friedman test yielded a $p-value$ of $0.0000001581$ $(< 0.05)$, which strongly suggests that there are statistically significant differences in harmfulness scores among the anti-patterns. Subsequently, the post-hoc Dunn test revealed that only two anti-patterns exhibited statistically significant differences compared to others: Golden Hammer differed from Hub-like Dependency, No CI/CD, Lack of Skeleton and Common Ownership; and Micro Frontend as the Goal differed from Hub-like Dependency. The implementation of the statistical tests and their results are available in our supplementary material (Section~\ref{sec:data}).

While Dunn's test did not indicate significant differences between most pairs, the boxplot visualization (Figure~\ref{fig:boxplots}) reveals notable trends. No CI/CD is perceived as significantly more harmful than other anti-patterns, as evidenced by its higher median score. Conversely, Golden Hammer is considered the least harmful, with the lowest median harmfulness score (6). These findings suggest that while statistical significance was not universally achieved, there is a strong perception that the absence of CI/CD is particularly harmful, whereas the reuse of familiar technologies is seen as acceptable.

\subsection{Thematic Analysis} \label{sec:qualitative} 

Based on the open coding quotations, the first author defined eight themes (Figure~\ref{fig:themes}). Subsequently, the second author independently categorized the same quotations using the established themes. On measuring the ICR, we obtained a Cohen's Kappa with a value of 0.84. According to Landis and Koch~\cite{landis1977measurement}, this is considered an almost perfect score, highlighting the reliability of our coding process and the robustness of the identified themes. In the following paragraphs, we present a brief explanation of each category and its main findings. The full thematic analysis is available in our supplementary material (Section~\ref{sec:data}).

\textbf{Commendations to the catalog} -- \textit{compliments to the catalog}: P2 acknowledged the novelty of the anti-patterns due to the lack of ones that focus on the MFE development, as stated in ``The term and technology is fairly new, and such patterns are not well established in the software community yet.'' P5 praised the catalog for addressing practical issues that practitioners face when working with MFE, as commented in ``This catalog highlights several real-world problems encountered in the day-to-day work with micro frontends.''

\textbf{How to use the catalog} -- \textit{how the catalog can be used to improve the development and maintenance of MFE architectures}: P2 highlighted that the catalog may provide essential guidance for best practices and avoidables in MFE development, as stated in ``It will serve as a guide for some `Dos' and `Donts' that are missing in the Micro Frontends world." P7 commented on the practical utility of the catalog in educating and integrating new team members in MFE software projects, as stated in ``Training new team members and onboarding them to micro frontend projects.'' P9 observed that the catalog may help in whether to use or not MFE architecture in ``think about architecture decisions and the decision to use micro frontends architecture or not.''

\textbf{Improvements to examples} -- \textit{proposals for new examples or enhancements to the presented ones}: For Cyclic Dependency, P4 suggests that the example should include all necessary dependencies to fully close the cycle, as noted, ``The example only states a dependency between A and B, but not another that would close the cycle between B and A.'' For Hub-like Dependency, P4 provided a practical scenario that emphasizes the importance of robust error handling in ``A main banking screen that has charts, lists, and balances. If this screen implements its own data fetch function that fails and renders the screen useless, then it is a problem.''

\textbf{Improvements to problem definition} -- \textit{enhancements to anti-patterns problems definitions}: P8 suggested the need for a clearer definition in Micro Frontends Greedy, once they had difficulty in distinguishing it from the Nano frontend or Mega frontends anti-patterns, as stated in ``I can't separate this anti-pattern from nano or mega frontends.'' For Common Ownership, P15 pointed out that small teams can also benefit from characteristics inherent of software modularization, as mentioned in ``Naturally, larger software involves more people, but I believe small teams can also benefit from software modularization, such as separation of layers and responsibilities, observability and maintainability.''

\textbf{Improvements to solutions} -- \textit{enhancements to anti-patterns solutions}: P2 suggested that the Knot Micro frontend anti-pattern should clearly define the difference between a good communication pattern and a bad one, as stated in  ``Solution could state what is a good communication pattern vs a bad one.'' For Hub-like Dependency, many participants suggested that the solution of avoiding aggregator screens does not address the problem correctly because, in the context of MFE architecture, screens will include many MFE fragments, as P2 emphasized in ``It is inevitable to have aggregators,'' and P20 in ``But not use it goes against the main idea of the MFE to be contextually segregated.'' P5 emphasized that the solution for Nano Frontend will rely on the organizational context it is inserted in, as stated, ``The solution here will depend entirely on the organizational context."

\textbf{Improvements to the catalog} -- \textit{enhancements to the catalog as a whole}: Most of the participants focused their suggestions for improving the catalog adding visual aids for enhancing the readability of the catalog, as P4 suggests in ``I would use some flow charts to exemplify most of the anti-patterns and make it more readable,'' and P18 complements in ``Add some diagrams and images help to understand some examples.''

\textbf{New anti-patterns} -- \textit{proposals of new anti-patterns}: P2 discussed the importance of choosing between build-time and runtime integration based on team and user needs, as stated in ``There are mainly two ways to integrate micro frontends: build-time and runtime. The decision on which to adhere reflects deeply in the teams' and users' needs more than the technical pros and cons each of them offer.'' P7 highlighted that data persistence on MFEs may become an anti–-pattern due to the issue when different MFE manage the state, as commented, ``Poor state management: Data persistence on MFEs when each frontend manages the state independently.'' Lastly, P8 suggests that Inconsistent User Experience is an existing issue in MFE.

\textbf{Proposal of new solutions} -- \textit{proposals of different solutions for specific anti-patterns}: For Mega Frontend, P4 suggested that better discussions between the product team and the development team could help define when features should be treated as different products, as stated in ``Lack of communication between the product team and the development team. It should be well discussed between the two teams to define when two or more features are different products.'' For Golden Hammer, P7 proposed adopting hybrid technology approaches for solving the problem, as commented in ``Adopting a hybrid technology approach supported by a common facade, such as a Backend for Frontend (BFF) layer, and using feature flags to manage gradual migration and experimentation might be a better approach.''

\section{Micro frontends anti-patterns}
In this section, we present the improved version of the MFE anti-patterns, including refinements, after analyzing practitioners' feedback. Due to the page constraint, we omitted the examples of the anti-pattern. These examples are available in our supplementary material (Section~\ref{sec:data}) and at our web application\footnote{\url{https://mfe-anti-patterns.online/micro-frontends-anti-patterns/\#/catalog}}.

\subsection{Cyclic Dependency}
\textbf{Category}: Inter-frontends

\textbf{Problem}: Two or more MFEs directly or indirectly depend on each other, resulting in high coupling between screens and fragments, compromising MFEs' independence and modularity. Thus, changes in one MFE require coordination with the others. Circular dependencies lead to challenges in a system's maintenance and evolution, compromising agility and the ability to scale developments efficiently.

\textbf{Solution}: High coupling between MFEs can be effectively mitigated through event-based communication, which removes the need for direct dependencies between MFEs. Instead, interactions are handled indirectly via a centralized event store. On implementing the Publish-Subscribe (Pub-Sub) pattern, an MFE can publish an event to the browser, allowing other MFEs to subscribe and respond when the event occurs. To ensure consistency and reduce errors, it is recommended to centralize event definitions in a shared library.

\subsection{Knot Micro Frontend}
\textbf{Category}: Inter-frontends

\textbf{Problem}: A Knot is composed of three or more MFEs whose communication with each other uses a context-specific interface. This means that navigation and data exchange between screens and fragments heavily depend on the unique context of each MFE involved. Adding new MFEs exacerbates the problem: as the number of MFEs grows, the interface complexity increases due to the introduction of new contexts, creating a highly coupled Knot that becomes difficult to maintain and integrate new functionalities.

\textbf{Solution}: A practical solution to address the problem of Knots is to implement domain-driven communication interfaces that are both generic and flexible. These interfaces should define a contract based on the domain model, specifying the essential fields required for each MFE to function correctly and interact with others. On designing new fields or attributes, it is essential to ensure their consistency and reusability and minimize tight coupling so other MFEs can utilize them. We recommend including a generic field in the interface containing a list of objects with standard properties such as label, value, and type, allowing each MFE to display data on the screen without needing to understand the specific meaning or context of the values. This approach reduces coupling between MFEs, while maintaining benefits such as modularity, scalability, and adaptability to new requirements.

\subsection{Hub-like Dependency}
\textbf{Category}: Inter-frontends

\textbf{Problem}: A screen of an MFE integrates fragments from several other MFEs, becoming a central point of interdependence. Any issue occurring in the main screen or one of its fragments can affect all other fragments present on it.

\textbf{Solution}: To prevent a single fragment failure from crashing the entire main screen, the screen should be kept as simple as possible, and each fragment should implement robust error-handling mechanisms. This can be achieved by implementing a strategy where uncaught errors within a fragment gracefully degrade its functionality, displaying a user-friendly fallback message. This approach ensures that users are informed of the issue without hindering their interaction with the remaining functionalities on the main screen.

\subsection{Nano Frontend}
\textbf{Category}: Intra-frontends

\textbf{Problem}: The frontend decomposes into numerous small MFEs with few screens or fragments. Small MFEs do not justify the cost of their maintenance. Furthermore, the presence of nano frontends can lead to issues of high coupling and the manifestation of other anti-patterns, such as cyclic dependency.

\textbf{Solution}: The issue of nano frontends arises when the definition of boundaries is inadequately and excessively granular. Adhering to Domain-driven Design~\cite{evans2004domain} principles is necessary to ensure an effective decomposition of MFEs. Therefore, the development team must work closely with the product team to gain a deep understanding of the domains and reflect them accurately in the architecture. To solve this issue, the architecture must be redesigned by grouping MFEs with the same domain is necessary. For minor variations within a domain, consider using templates or component libraries. This approach avoids creating a separate MFE for each slight variation, promoting efficiency and code reuse.

\subsection{Mega Frontend}
\textbf{Category}: Intra-frontends

\textbf{Problem}: Decomposing the architecture into a few MFEs encompassing numerous screens and fragments manifests this anti-pattern. The MFE inherits the challenges of a monolithic frontend, such as difficulties in testing, slow builds and deployments, high coupling between its components, lack of modularity, and limited scalability.

\textbf{Solution}: To avoid this problem, the development team must work closely with the product team to gain a deep understanding of the domains and reflect them accurately in the architecture. To fix this issue, the team should reevaluate the architecture and divide the MFEs into more granular units, separating functionalities into smaller and specialized MFEs based on domains. This approach helps reduce complexity, enhance maintainability, and foster a modular and scalable architecture.

\subsection{Micro Frontend Greedy}
\textbf{Category}: Intra-frontends

\textbf{Problem}: When a developer is uncertain about creating a new MFE, the common practice is to opt for its creation. Whenever a need arises to develop a new set of screens or fragments, a new MFE is instantiated. This can lead to an explosion in the number of MFEs, making the system difficult to understand and increasing the likelihood of both nano and mega frontends emerging.

\textbf{Solution}: To determine where to implement a new feature composed of a set of screens and/or fragments, the domain of the new feature must first be defined. If it falls within the domain of an existing MFE, it should be implemented there. In this case, a summary of all MFEs, their contexts, and domains can help identify the best fit for the new feature. If it belongs to a brand new domain, one or more MFEs should be defined based on the domain definition. Establishing well-defined domains relies on collaboration between the development and product teams to define boundaries accurately.

\subsection{No CI/CD}
\textbf{Category}: Operations

\textbf{Problem}: The company lacks an automated Continuous Integration (CI) and Continuous Delivery (CD) pipeline, so developers must manually execute tests and perform deployments. This manual process becomes burdensome, especially with the potential existence of multiple MFEs. It increases development time, reduces productivity, and raises the risk of errors in the production environment.

\textbf{Solution}: Implement an automated and replicable CI/CD process that extends for new MFEs, ensuring they will have automated test execution and deployment consistently and efficiently. This should be part of the Definition of Done (DoD) of the architecture.

\subsection{No Versioning}
\textbf{Category}: Operations

\textbf{Problem}: The MFEs are not versioned. Small and large changes can impact the integration between different MFEs and cause errors. Consequently, the MFEs become less independent, requiring coordinated deployments.

\textbf{Solution}: Adopting a versioning approach like Semantic Versioning is essential to ensure that changes do not impact functioning versions. For example, consider a fragment that is used in screens across different MFEs in a client-side rendering scenario. Without versioning, any change to the fragment's parameters or return values could break the interaction on all the screens it integrates with. However, with versioning, such updates would not impact the current versions used by other MFEs, as they can continue to request the previous version of the fragment and update at their convenience. This approach helps maintain a stable environment and minimizes disruptions caused by updates.

\subsection{Lack of skeleton}
\textbf{Category}: Operations

\textbf{Problem}: No skeleton or predefined boilerplate is available as a base for creating new MFEs. This leads to the creation of MFEs from scratch or based on an existing MFE and inheriting its issues. The consequences include wasted time, increased risk of errors, duplicated code across MFEs, and a need for more standardization in development.

\textbf{Solution}: Whenever a new technology is used to implement an MFE, the development team must create a repository containing the necessary base code, known as a boilerplate. The boilerplate should enable the creation of new MFEs with the same technology by simply cloning it. Keeping the boilerplate updated with new design patterns and library versions is crucial. Additionally, the development team should create comprehensive documentation detailing the entire process of creating a new MFE, regardless of the technology. This documentation should provide instructions on adding automated CI/CD, integrating the MFE into the existing system, and addressing other relevant aspects.

\subsection{Common Ownership}
\textbf{Category}: Development 

\textbf{Problem}: A single team is tasked with managing all MFEs, which can occur either due to a lack of team division or when teams are segmented based on technical aspects such as data, frontend, and backend. However, one of the key benefits of MFE architecture is independence, so adopting MFE Architecture without distinct teams to operate independently negates this advantage.

\textbf{Solution}: Context should be the defining factor when structuring development teams. Therefore, defining the boundaries of teams and MFEs is essential according to Domain-driven Design~\cite{evans2004domain}, so a team will be responsible only for MFEs within its domain. Creating shared libraries can facilitate boundary definition and promote greater team independence.

\subsection{Golden Hammer}
\textbf{Category}: Development 

\textbf{Problem}: All MFEs utilize the same technology, even if it does not meet the specific needs of each MFE. It happens due to developers' familiarity with only one specific technology. This approach limits the architecture, failing to take advantage of the benefits of the possibility of a heterogeneous architecture, which is one of the main attractions of adopting MFEs.

\textbf{Solution}: To choose the most suitable technology that addresses the specific challenges of each MFE, which includes adopting the correct programming languages, frameworks, and libraries during its development. When uncertain about a particular technology, conducting a proof-of-concept (POC) can validate its suitability. Testing new technologies through POCs helps validate their suitability without compromising the establishment of standardized patterns within the company. However, it's important to note that increasing the variety of technologies can increase the complexity of the architecture.

\subsection{Micro Frontend as the goal}
\textbf{Category}: Development 

\textbf{Problem}: Adopting the MFE architecture in inappropriate contexts can lead to more issues than benefits, especially in systems with few screens and low complexity or in companies lacking a sufficient number of developers to create dedicated teams for different application domains. In such situations, the maintenance costs of the architecture may outweigh the expected benefits, making its implementation unfeasible.

\textbf{Solution}: Software teams must consider carefully different aspects of adopting MFE architecture. Considering the system's complexity, the feasibility of maintaining automated CI/CD pipelines and the team's restructuring according to different domains is necessary.

\section{Discussion}
The proposed MFE anti-patterns are closely aligned with their MS counterparts, reflecting the inherent similarities between these architectural styles. Given the frequent evolution of software systems from monolithic architectures to MS and subsequently to MFE, anti-patterns related to development and operation like No CI/CD and Common Ownership are likely to persist in MFE if they were previously encountered in MS.

We observed that the proposed anti-patterns have varying impacts on developers and end users. The No Versioning and Hub-like Dependency anti-patterns significantly affect end users, potentially causing application crashes. The Golden Hammer anti-pattern has a moderate impact on both end users and developers, stemming from poor experiences due to inappropriate technology choices. The remaining anti-patterns primarily impact developers, complicating architecture maintenance and evolution, though they have a low direct impact on end users.

By accessing our online catalog, developers can learn how to avoid bad practices when working with MFE from an organizational and architectural point of view. The catalog can act as a checklist or as a management resource for experienced or new members of software projects. Moreover, the catalog may also help practitioners recognize bad practices that have become standardized within their organizations due to their routine use and familiarity. Anecdotal evidence suggests that the catalog is already valuable to developers. For instance, some survey participants reported using it in their daily work, as P12 mentioned: ``Every time I have to implement a new feature on a micro frontend, I consult the catalog to remember the anti-patterns.'' This feedback highlights the catalog's role in enhancing development practices and fostering awareness of best practices in MFE design.

While this study has revealed a strong correlation between MS and MFE anti-patterns, there remain specific anti-patterns unique to MFE architectures that warrant further exploration. Issues related to UI inconsistency, the management of the state through global versus local stores, and the selection of inappropriate composition approaches have not yet been addressed by the proposed anti-patterns. Future research should focus on identifying and mitigating these and other MFE-specific anti-patterns to enhance the overall quality and effectiveness of MFE architectures.

To ensure high-quality system design and prevent software degradation, it is crucial to identify anti-patterns early and perform the necessary refactoring. Automating the detection of these anti-patterns may allow for early intervention, thus significantly mitigating their impact on software projects. Therefore, future research should also prioritize the development of automated detection methods tailored to MFE architectures. Such advancements will not only improve system quality but also help prevent long-term negative consequences.

Lastly, it is also important to address new anti-patterns that occur during software development. As presented in the thematic analysis result, participants proposed new anti-patterns related to Inconsistent User Experience, Fragmented State Management, Complex Inter-MFE Communication, Overhead of Independent Deployments, Security and Authentication Challenges, Performance Bottlenecks, Security Risks and Observability. These anti-patterns' problems and solutions must be clearly defined and validated by practitioners. It highlights the need for ongoing research to address emerging challenges in the MFE field, ensuring the development of efficient MFE architectures.

\section{Threats to validity}

\textbf{Internal Validity.} (1) The length of the form used to gather practitioner's feedback. A long form may fatigue the participant, affecting their responses. To address this issue, we provided an estimated completion time to participants when inviting them and structured the form to present each anti-pattern on a separate page, allowing participants to focus on one anti-pattern at a time. We also included a progress bar to give participants a clear indication of how many anti-patterns remained to be evaluated. (2) Participants representativeness. To address this, we focused on inviting participants with previous backgrounds in working with MFE in the industry. To ensure that participants were not biased by the results of previous works, we did not propose a predefined set of bad practices to the participants.

\textbf{External Validity.} Subjects sample size. To address this threat, we distributed the survey to a broader pool of software engineers, encompassing frontend developers, fullstack developers, and team leaders. It allowed us to capture a broader range of perspectives from practitioners with varying levels of MFE knowledge and involvement. While the sample size may not be ideal for full population generalization, the diversity of participant roles strengthens the applicability of our findings to real-world MFE development practices.

\textbf{Conclusion Validity.} (1) One might reach incorrect conclusions given the data. On addressing it, multiple researchers were involved in the data interpretation. For both the quantitative and thematic analysis, two researchers handled the data interpretation and categorization of themes, and a third, who is an expert in Software Engineering with more than 20 years of experience, reviewed all the results. (2) The harmfulness scores calculated using the median may not fully capture the nuances of participants' perceptions. Although we do not provide enough evidence on the harmfulness of ranking the anti-patterns, we illustrate which of them the practitioners consider the most harmful during software development. We also emphasize that our survey focuses on validating the initial anti-patterns, instead of generating the ranking itself.

\textbf{Construct Validity.} Influence of researcher bias on the qualitative results. To mitigate it, we employed a two-coder approach to the thematic analysis, with the second author independently reviewing the qualitative data and conducting a separate thematic analysis. Following this, we employed Cohen's Kappa~\cite{cohen1960coefficient} to measure the level of agreement between the two coders. The resulting score of 0.84, classified as ``excellent agreement,'' strengthens our confidence in the objectivity and trustworthiness of the identified themes.

\section{Related work}
Although some practitioners have shared their experience on anti-patterns in MFE applications on blogs and keynotes \cite{mezzalira2023microfrontends,shinde20224,rappl2024top,geeks2024micro}, no scientific studies have been conducted to propose a catalog of MFE anti-patterns. We conducted a Rapid Review -- available at our supplementary material (Section~\ref{sec:data}) -- and found no studies proposing anti-patterns for MFEs. Mezzalira \cite{mezzalira2021building} briefly mentions that sharing any state between micro-frontends is considered an anti-pattern but does not provide an in-depth definition or exploration of this or other anti-patterns. Peltonen et al. \cite{peltonen2021motivations} conducted a Multivocal Literature Review and stated that researchers have not yet deeply investigated MFEs, and patterns and anti-patterns have not been defined. Taibi et al. \cite{taibi2022micro} presented a set of development approaches based on their experience and reported the lack of pattern and anti-pattern definitions. Moraes et al. \cite{moraes2024micro} reported a case study findings, which revealed that involuntary anti-patterns may occur since they are not yet mature, generating severe negative impacts on software projects.

Several papers present case studies or experience reports on implementing MFE architectures. M{\"a}nnist{\"o} et al. \cite{mannisto2023experiences} presented their experience through the migration of a monolithic frontend to the MFE architecture. Capdepon et al. \cite{capdepon2023migration} proposed an approach to migrate from monolithic mobile architecture to MFE. Moraes et al. \cite{moraes2024micro} provided an experience report demonstrating how the same application can be implemented using different MFE approaches. Perlin et al. \cite{perlin2023approach} presented a case study of how to implement an MFE application with Webpack. Kaushik et al. \cite{kaushik2024micro} proposed a framework for the design of web applications with MFEs and microservices. They conducted a case study to empirically analyze and evaluate the effectiveness of the proposed framework. Pavlenko et al. \cite{pavlenko2020micro} implemented an MFE case study to report and discuss the issues arising during development. All these papers can assist in making architectural decisions when implementing an MFE architecture. However, they cannot be used to evaluate an architecture or serve as a guide for identifying hidden problems within it.

Given the shared characteristics of microservices and MFE architectural styles, a study of microservices anti-pattern was undertaken to propose the MFE anti-patterns. Through a Systematic Literature Review (SLR), Cerny et al.~\cite{cerny2023catalog} crafted a catalog with 58 disjoint microservice anti-patterns grouped into five categories: Intra-service, Inter-service, Service interaction, Security, and Team Anti-patterns. Tighilt et al.~\cite{tighilt2020study} proposed a catalog with 16 microservice anti-patterns based on an SLR and the analysis of 67 systems, examining them for potential violations of microservices principles and design practices that could be indicative of anti-patterns. Building on practitioner experience, Taibi et al.~\cite{taibi2020microservices} introduced a catalog and a taxonomy of the most common microservices anti-patterns. Their three-year interview study identified 20 anti-patterns, including organizational and technical anti-patterns. Bogner et al.~\cite{bogner2019towards} conducted an SLR to propose a taxonomy of 36 microservices anti-patterns. Additionally, they developed a collaborative web application that allows users to explore and interact with their catalog. While both architectural styles share common anti-patterns, a dedicated catalog for MFEs is necessary to identify and address these issues specifically within the MFEs context.

\section{Conclusions}
This paper presents a catalog of 12 MFE anti-patterns derived from microservices anti-patterns and the analysis of real-world problems faced by MFE development teams. To validate whether the identified problems are prevalent in MFE architectures and if the proposed solutions address them effectively, we conducted a survey with 20 industry practitioners. Based on their feedback, we improved the anti-patterns and showcased them in a web application to foster community collaboration.

The survey results show that all identified anti-patterns have been encountered by participants in real-world MFE projects, each receiving varying harmfulness scores. Participants emphasized the catalog's utility as a valuable resource for improving MFE architecture, highlighting its potential to guide both novice and experienced developers in avoiding common pitfalls. Additionally, participants provided valuable insights by suggesting new anti-patterns. This highlights the importance of creating a collaborative platform where researchers and practitioners can jointly propose, discuss, and validate new anti-patterns. Our web application facilitates this process by enabling practitioners to share their knowledge and experiences, ensuring the catalog remains comprehensive and up-to-date.

For future work, we plan to expand the anti-patterns catalog by: (1) analyzing the suggestions the participants provided through the survey; (2) reviewing published MFE architectures to identify recurring issues that deviate from MFE guidelines; (3) conducting case studies in industrial settings to uncover common problems in real MFE architectures; (4) using our collaborative catalog on GitHub to continuously collect feedback from practitioners, helping to identify new anti-patterns based on real-world experiences. We intend for industry practitioners to assess all newly proposed anti-patterns.

We also aim to develop automated detection tools using Dependency Graph and Static Analysis tools to identify anti-patterns automatically, integrate these tools into code review systems, and design visualization tools to aid in detecting anti-patterns in MFE architectures. For the web application, we plan to use GitHub Insights and GitHub Discussions to establish contributor leaderboards, helping to motivate developers to engage more actively and contribute to the catalog. 

\section{Data availability}\label{sec:data}
The initial and improved version of the anti-patterns, including examples, survey responses, and the qualitative and quantitative analysis conducted in this research, are currently maintained as an open-source project accessible at: \url{https://zenodo.org/records/14868084}.

\section*{Acknowledgment}
We thank all the participants in the empirical study and USES Research Group members for their support. The present work is the result of the Research and Development (R\&D) project 001/2020, signed with the Federal University of Amazonas and FAEPI, Brazil, which has funding from Samsung, using resources from the Informatics Law for the Western Amazon (Federal Law nº 8.387/1991), and its dissemination is in accordance with article 39 of Decree No. 10.521/2020. This work was carried out with the support of the Coordination for the Improvement of Higher Education Personnel - Brazil (CAPES-PROEX) -- Funding Code 001;  CNPq process 314797/2023-8; CNPq process 443934/2023-1; and CNPq process 445029/2024-2. Additionally, this work was partially funded by the Foundation for Research Support of the State of Amazonas -- FAPEAM -- through the PDPG project.

\bibliographystyle{IEEEtran}
\bibliography{IEEEabrv,references}

\end{document}